%% file: main.tex
\documentclass[9pt,shortpaper,twoside,web]{ieeecolor}
\usepackage{generic}
\usepackage{cite}
\usepackage{amsmath,amssymb,amsfonts}
\usepackage{algorithmic}
\usepackage{graphicx}
\usepackage{textcomp} 
\usepackage{tabularx}
\usepackage{array}
\usepackage{subcaption}
\usepackage{xcolor}
\usepackage{multirow}
\usepackage{colortbl}
\usepackage{makecell}
\usepackage{enumerate}
\usepackage[outdir=./]{epstopdf}
\def\BibTeX{{\rm B\kern-.05em{\sc i\kern-.025em b}\kern-.08em
    T\kern-.1667em\lower.7ex\hbox{E}\kern-.125emX}}
\begin{document}
\title{EEG-Derived Voice Signature for Attended Speaker Detection}
\author{Hongxu Zhu, \IEEEmembership{Member, IEEE}, Siqi Cai, \IEEEmembership{Member, IEEE}, Yidi Jiang, \IEEEmembership{Student Member, IEEE}, \\ Qiquan Zhang, \IEEEmembership{Member, IEEE}, and Haizhou Li, \IEEEmembership{Fellow, IEEE}
\thanks{Manuscript received  ; revised . 
The work is funded by National Natural Science Foundation of China (Grant No. 62271432); Internal Project of Shenzhen Research Institute of Big Data under grant No. T00120220002; Guangdong Provincial Key Laboratory of Big Data Computing, The Chinese University of Hong Kong, Shenzhen (CUHK-Shenzhen), China (Grant No. B10120210117-KP02); Human-Robot Collaborative AI for Advanced Manufacturing and Engineering (Grant No. A18A2b0046), Agency for Science, Technology and Research, Singapore; the Deutsche Forschungsgemeinschaft (DFG) under Germany's Excellence Strategy (University Allowance, EXC 2077, University of Bremen, Germany).}
\thanks{Hongxu Zhu, Siqi Cai, Yidi Jiang, Qiquan Zhang, and Haizhou Li are with the Department of Electrical and Computer Engineering, National University of Singapore, Singapore.}
\thanks{Qiquan Zhang is also with the School of Electrical Engineering and Telecommunications, University of New South Wales, Sydney, NSW 2052, Australia. 
}
\thanks{Haizhou Li is also with Shenzhen Research Institute of Big Data, the School of Data Science, The Chinese University of Hong Kong, Shenzhen (CUHK-Shenzhen), China, Kriston AI Lab, China, and the Machine Listening Lab, University of Bremen, Germany.}
\thanks{Corresponding author: Siqi Cai (e-mail: elesiqi@nus.edu.sg).}}

\maketitle

\begin{abstract}
\textit{Objective:} Conventional EEG-based auditory attention detection (AAD) is achieved by comparing the time-varying speech stimuli and the elicited EEG signals. However, in order to obtain reliable correlation values, these methods necessitate a long decision window, resulting in a long detection latency. Humans have a remarkable ability to recognize and follow a known speaker, regardless of the spoken content. In this paper, we seek to detect the attended speaker among the pre-enrolled speakers from the elicited EEG signals. In this manner, we avoid relying on the speech stimuli for AAD at run-time. In doing so, we propose a novel EEG-based attended speaker detection (E-ASD) task. \textit{Methods:} We encode a speaker's voice with a fixed dimensional vector, known as speaker embedding, and project it to an audio-derived voice signature, which characterizes the speaker's unique voice regardless of the spoken content. We hypothesize that such a voice signature also exists in the listener's brain that can be decoded from the elicited EEG signals, referred to as EEG-derived voice signature. By comparing the audio-derived voice signature and the EEG-derived voice signature, we are able to effectively detect the attended speaker in the listening brain. \textit{Results:} Experiments show that E-ASD can effectively detect the attended speaker from the 0.5s EEG decision windows, achieving 99.78\% AAD accuracy, 99.94\% AUC, and 0.27\% EER. \textit{Conclusion:} We conclude that it is possible to derive the attended speaker's voice signature from the EEG signals so as to detect the attended speaker in a listening brain. \textit{Significance:} We present the first proof of concept for detecting the attended speaker from the elicited EEG signals in a cocktail party environment. The successful implementation of E-ASD marks a non-trivial, but crucial step towards smart hearing aids.
\end{abstract}


\section{Introduction}
\label{sec:introduction}

With smart hearing aids, we improve the speech perception of the hearing-impaired by enhancing the desired voice while suppressing background noise. However, in cocktail party problems (CPP) where multiple speakers are talking at the same time~\cite{cherry1953some}, the hearing aids are prone to malfunction. This is due to the difficulty in singling out the desired sound source. In fact, reliably identifying the attended voice solely from speech signals is challenging. Humans have a remarkable ability to follow attended speaker in CPP, as the ears form a close loop with the brain, but the current hearing aids don't. To overcome the limitations of hearing aids, a potential solution is to incorporate neurosteering techniques that utilize brain signals to extract attention-related information, specifically known as auditory attention detection (AAD)~\cite{mcdermott2009cocktail,geirnaert2021electroencephalography}. 

In the last decade, AAD has been achieved with various brain signals, such as electrocorticography (ECoG)~\cite{mesgarani2012selective}, magnetoencephalography (MEG)~\cite{ding2012neural,akram2016dynamic}, and electroencephalography (EEG)~\cite{o2015attentional, biesmans2016auditory, crosse2016multivariate,wong2018comparison, de2020machine, thornton2022robust, xu2022decoding, de2018decoding, monesi2020lstm, accou2021modeling, geirnaert2020fast, vandecappelle2021eeg, su2022stanet, Deckers475673, kuruvila2021extracting, 9633231}. From a practical point of view,  non-invasive and wearable EEG devise is a more feasible option. Prior EEG-based AAD solutions can be mainly grouped into three categories according to the reference information as: stimulus-based~\cite{o2015attentional, biesmans2016auditory, crosse2016multivariate,wong2018comparison, de2020machine, thornton2022robust, xu2022decoding, de2018decoding, monesi2020lstm, accou2021modeling }, locus-based~\cite{geirnaert2020fast, vandecappelle2021eeg, su2022stanet}, and directional-stimulus-based schemes~\cite{Deckers475673, kuruvila2021extracting, 9633231}. 

The exploration of EEG-based AAD initially focuses on stimulus-based approaches, which aim to establish a relationship between EEG signals and the attended speech stimulus. The classic stimulus-reconstruction
approach estimates the linear\cite{o2015attentional, biesmans2016auditory, crosse2016multivariate,wong2018comparison} or non-linear \cite{de2020machine, thornton2022robust, xu2022decoding} mapping between the attended speech stimulus (mainly speech envelope) and EEG responses. With the advent of deep learning techniques, match-mismatch (MM) approaches \cite{de2018decoding, monesi2020lstm, accou2021modeling} are further proposed to predict the match probability between a pair of speech stimuli and EEG responses without the need for stimulus reconstruction. 
As brain activity also reflects the location of auditory attention~\cite{wostmann2016spatiotemporal}, the locus-based AAD approaches attempt to detect the directional focus of attention other than the attended speaker. These models, referred to as spatial AAD, linearly~\cite{geirnaert2020fast} or non-linearly \cite{vandecappelle2021eeg, su2022stanet} predict the direction of attention (e.g., left or right) from the EEG inputs. Because the spatial AAD models don't require the availability of speech stimulus, they can function well in noisy or cluttered environments. 
The directional-stimulus-based approaches exploit both the direction information and speech stimulus to determine attention. The common operation of this method~\cite{Deckers475673, kuruvila2021extracting, 9633231} is concatenating the EEG signal with the two competing stimuli for direct classification. The order of the speech stimulus needs to be pre-determined to indicate the directions of the two speakers. For instance, the left/right stimulus could be added as the top/bottom channels to EEG signals~\cite{Deckers475673}. Hence, if the permutations of the competing speakers are wrongly presented, the classification performance could be significantly dropped.

Despite much progress, the locus-based or directional-stimulus-based AAD approaches might only have a joint action with a microphone array that can also decode the direction information\cite{aroudi2020cognitive}. 
Single microphone schemes lack this capability, however, they hold greater application potential for low hardware requirements. In contrast, stimulus-based approaches could be combined with single microphone schemes \cite{das2020linear,ceolini2020brain}, but they typically have lower detection accuracy than the other two types of solutions. Prior works \cite{o2015attentional, biesmans2016auditory, crosse2016multivariate,wong2018comparison, de2020machine, thornton2022robust, xu2022decoding, de2018decoding, monesi2020lstm, accou2021modeling} mainly focus on estimating the attention to the time-varying speech features, e.g., envelope and spectrogram, which is always challenging due to the noisy nature of EEG signals. 

\begin{figure}[!t]
\centerline{\includegraphics[width=0.8\columnwidth]{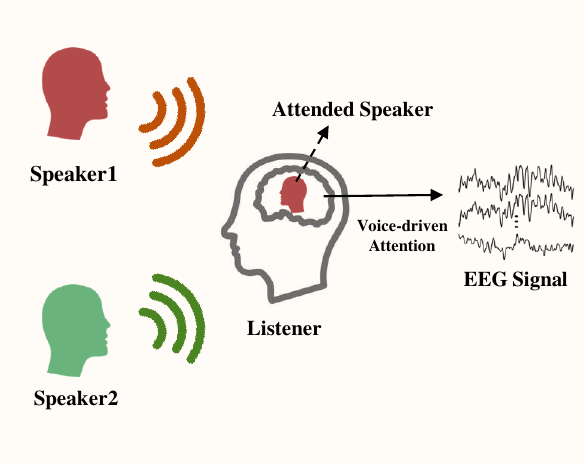}}
\caption{The voice-driven auditory attention. Humans have the ability to attend to a particular speaker easily in a complex multi-talker acoustic environment, regardless of the spoken content. The speaker identity guides the top-down attention to the voluntary allocation of neural resources. We hypothesize that the voice of the attended speaker elicits the listener’s EEG signals.} 

\label{voice_attention}
\end{figure}

Similar noise issues are quite common in wireless communication systems. To alleviate the noise problem, the time-varying information is modulated with a time-invariant symbol set (e.g., quadrature amplitude modulations (QAM) \cite{zhu2016ber}) for transmission and decoding. By simplifying the decoding problem from `estimating the time-varying information' to `detecting the best-matched symbols from a symbol set', high decoding performance (bit error rate) can still be achieved even in low signal-to-noise-ratio (SNR) scenarios. 
Indeed, in the domain of EEG-based AAD, the locus-based methods \cite{geirnaert2020fast, vandecappelle2021eeg, su2022stanet} also follow this strategy (using directions as symbols), which is one of the fundamental reasons for their success. In the same spirit, if we can `modulate' the speech stimulus with a time-invariant symbol set and then detect these symbols from the time-varying EEG signals, the noise problem from the EEG signal can also be alleviated. Fortunately, the speaker's \textit{distinct voice} is a natural symbol set for speech signals. 
With a unique vocal construct and speaking manner, e.g., vocal tract length, larynx size, accent, and prosody, each speaker features unique personal traits~\cite{kinnunen2010overview}. Furthermore, humans have a remarkable ability to recognize the voice of a known speaker, regardless of the speech content. Recently, neuroscience research has also shown that humans are more easily to attend to a known person\cite{peelle2022our}. In the case of attending to an unfamiliar speaker, listeners usually adjust to this speaker over time and become more efficient at understanding their speech. This is because the human brain is portrayed as a ‘prediction machine’ where top-down expectations are constantly predicting bottom-up information. As shown in Fig. \ref{voice_attention}, the top-down attention, which is established on cognitive factors (e.g., speaker identity, knowledge of speech, and language), guides the voluntary allocation of neural resources \cite{blank2016prediction}. 
As demonstrated in \cite{hu2021speaker}, EEG signals elicited by different speakers have discriminative patterns, which indicated the important role of voice information in top-down attention. However, the previous stimulus-based AAD studies \cite{o2015attentional, biesmans2016auditory, crosse2016multivariate,wong2018comparison, de2020machine, thornton2022robust, xu2022decoding, de2018decoding, monesi2020lstm, accou2021modeling}, which detects the attended stimulus, are mostly driven by the bottom-up attention. 
The AAD solution based on top-down attention has not yet been explored.

In this paper, we mimic the voice-driven top-down attention and aim to detect the attended speaker from the EEG signals. 
The deep speaker embedding techniques \cite{desplanques2020ecapa, bai2021speaker} could feature the speaker's voice into a vector, namely speaker embedding, from pre-enrolled short utterances by emphasizing speaker-specific properties while suppressing speech content.
The speaker embedding is robust to represent a speaker's voice and has been widely adopted in other speech processing tasks, such as speaker extraction~\cite{jiang2023target}. 
In this paper, we propose a novel EEG-based AAD task, i.e. E-ASD, that directly detects the attended speaker identity from the time-varying EEG signals. The main contributions of this study are four-fold,

\begin{itemize}
    \item We propose a novel AAD task based on top-down attention that detects the identity of the attended speaker from the elicited EEG signals. To the best of our knowledge, this is the first of its kind in the AAD domain. 
    \item We investigate the feasibility of leveraging pre-enrolled speaker embeddings to complement the EEG signals in detecting the voice signature of an attended speaker.
    \item We propose E-ASD, which estimates the matching likelihood between a window of EEG signals with a pre-enrolled speaker embedding. To determine auditory attention, the speaker whose embedding induces the highest matching value is selected.
    \item We conduct comprehensive experiments to evaluate the performance of the E-ASD. The results show that the E-ASD model achieves robust decoding performance, which is a non-trivial, but crucial step for the neurosteered speech extraction task.
\end{itemize}

The rest of the paper is organized as follows. In section \ref{sec:2}, we present the diagram of the E-ASD and the technical details of each part. The experimental setup is shown in Section \ref{sec:3} whereas the experiment findings are reported in Section \ref{sec:4}. In Section \ref{sec:5}, we discuss these results and consider future works. In section \ref{sec:6}, we conclude this paper.

\section{EEG-based Attended Speaker Detection}
\label{sec:2}

\begin{figure*}[!t]
\centerline{\includegraphics[width=2\columnwidth]{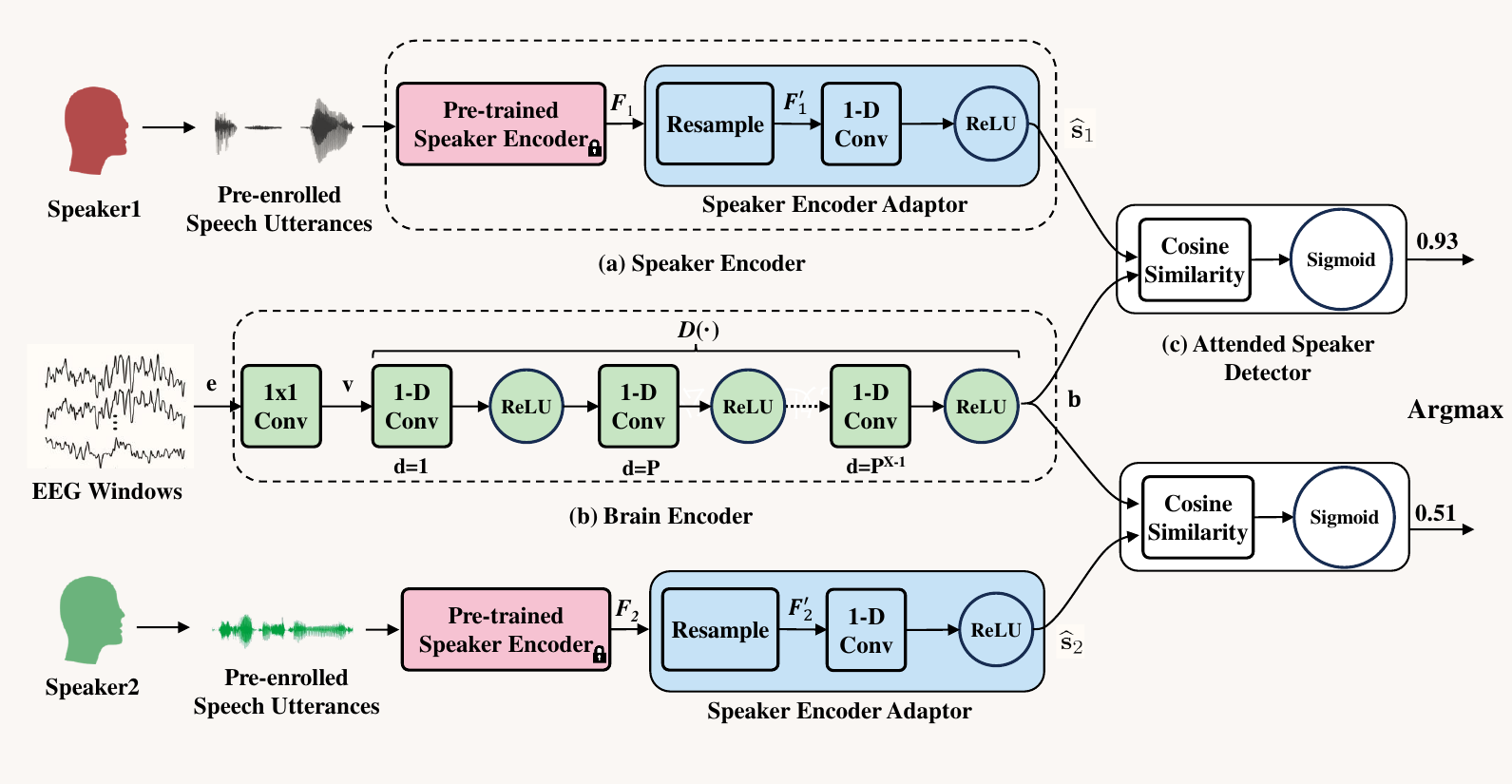}}
\caption{\textcolor{black}{The EEG-based Attended Speaker Detection network (E-ASD): (a) a pretrained speaker encoder encodes one or more of the pre-enrolled speech utterances into a fixed dimension speaker embedding, i.e. $F_1$ or $F_2$. An adaptor further projects the speaker embedding into an audio-derived voice signature, i.e. $\widehat{\mathbf{s}}_1$ or $\widehat{\mathbf{s}}_2$;
(b) a brain encoder decodes an EEG-derived voice signature $\mathbf{b}$ from an EEG decision window $\mathbf{e}$; and (c) an attended speaker detector estimates the matching likelihoods across multiple possible pairs of audio-derived and EEG-derived voice signatures $[\widehat{\mathbf{s}}_j, \mathbf{b}]$. 
The highest matching likelihood represents the detected attention.
The E-ASD model is trained to associate the EEG signals with the voice signature of the attended speaker.
The audio-informed voice signature is derived from pre-enrolled speech utterances, which could be pre-registered into the system. Therefore, E-ASD could detect the attended speaker without the speech stimuli at run-time.
(Note: a lock denotes a pretrained model with frozen parameters, that are not involved in E-ASD training.)} 
}
 
\label{Diagram}
\end{figure*}


The EEG-based attended speaker detection aims to detect the attended speaker identity from the time-varying EEG signals. 
Here, the E-ASD network consists of three main processing modules, as shown in Fig.~\ref{Diagram}: (a) speaker encoder, (b) brain encoder, and (c) attended speaker detector. 

A speaker encoder is employed to characterize a speaker by encoding one or more of his/her spoken utterances into an audio-derived voice signature.
A brain encoder is employed to decode the attended speaker's voice signature from the elicited EEG signal, that is called EEG-derived voice signature.
Finally, an attended speaker detector estimates the matching likelihood across multiple possible pairs of audio-derived and EEG-derived voice signatures to detect the attended speaker.
 

To perform attended speaker detection, the matching likelihood for each speaker is predicted and the attention is detected as the one with the highest matching value. For clarity, we only present a two-speaker scenario. However, it should be noted that E-ASD can be easily generalized to multiple speakers. We first formulate the problem and describe the details of each stage in this section.

\subsection{Problem Formulation}
\label{system model}
In a cocktail party environment, suppose that we have $C$ speakers, denoted as $s_1, \ldots, s_c$, talking simultaneously. In this paper, we discuss the case where $C=2$. The listener attends to the $j$-th speakers $s_j$ and elicits the EEG signal $e(t)$. We hypothesize that the listener forms a voice signature $\mathbf{b}$, that is called EEG-derived voice signature, in the listener's brain and guides the top-down attention. In other words, the EEG signal $e(t)$ reflects the attended speaker $\mathbf{b}$ as:


\begin{equation}
e(t) = B\left(\mathbf{b}\right) 
\end{equation}
where 
$B(\cdot)$ denotes the elicited EEG signals in response to the speech by the attended speaker. Its inverse function, $\mathbf{b} = B^{-1}(e(t))$, represents the process of decoding the voice signature from the EEG signals, that is called EEG-derived voice signature.

In this study, we assume that we have the facility to pre-enroll the two speakers' voices in advance to obtain their audio-derived voice signatures, i.e. $\widehat{\mathbf{s}}_1$ and $\widehat{\mathbf{s}}_2$. This is called speaker enrollment. At inference time, we don't need to acquire the audio-derived voice signature from the speech stimuli. Instead, we use a pre-enrolled audio-derived voice signature as the reference to detect in the elicited EEG signals whether it is being attended by a listening subject. In practice, we are comparing the pre-enrolled audio-derived voice signature of a speaker and the EEG-derived voice signature. The former is encoded by the speaker encoder, and the latter is encoded by the brain encoder. 
\begin{equation}
j = \underset{j \in 1, 2}{\operatorname{argMax}} \ f\left(\widehat{\mathbf{s}}_j, \mathbf{b} \right) 
\end{equation}
where $f(\cdot)$ denotes a likelihood estimation function to predict the matching probability.

\subsection{Speaker Encoder}
\label{Speaker_encoder}
The speaker encoder, as depicted in Fig. \ref{Diagram} (a), seeks to characterize the unique voice signature of a speaker with the given spoken utterances, regardless of the speech content. This has been a long-standing study in speaker verification \cite{desplanques2020ecapa, bai2021speaker}.
A general speaker verification pipeline consists of two modules: (a) a front-end that transforms a spoken utterance into a fixed-dimension speaker embedding. (b) a back-end that measures the similarity between the speaker embeddings of an enrollment and a test speech sample~\cite{bai2021speaker}. Usually, a neural network based front-end is trained on a speech corpus on thousands of speakers. 
It then serves as the speaker encoder to transform a spoken utterance into a speaker embedding for speaker comparison during training and testing. 

It is out of the scope of this paper to study the training of the speaker encoder. We adopt a publicly available speaker encoder and freeze its parameters. We further use a limited amount of speech data available in the EEG-Speech dataset to train an adaptor to serve two purposes. One is to adapt the general purpose speaker encoder to the task of attended speaker detection. Another is to project the speaker encoder output, i.e. speaker embedding, to a vector space that has the same dimension as the EEG-derived voice signature. 




\textcolor{black}{For $K$ enrollment utterances of the $i$-th speaker, each utterance is encoded into a $L$-dimension speaker embedding. We use the average of $K$ such speaker embeddings, $\mathbf{F}_i \in \mathbb{R}^{1 \times L}$, to characterize the $i$-th speaker. For brevity, we drop the speaker index $i$ and have the speaker embedding $\mathbf{F}$. To facilitate the comparison between the audio-derived voice signature $\widehat{\mathbf{s}}$, and the EEG-derived voice signature $\mathbf{b} \in \!\mathbb{R}^{1 \times S}$, of which the dimension is related to the EEG window size $S$, we employ a speaker encoder adaptor that resamples the speaker embedding $\mathbf{F} \in \mathbb{R}^{1 \times L}$ to $\mathbf{F}^{\prime}\!\in\!\mathbb{R}^{1 \times S}$, to derive the voice signature $\widehat{\mathbf{s}}=\operatorname{ReLU}(\mathrm{Conv1D}(\mathbf{F}^{\prime}))$. In this way, we adapt the speaker embedding $\mathbf{F}$ to the space of $\mathbf{b} \in \!\mathbb{R}^{1 \times S^{\prime}}$. It is noted that typically the dimension of a speaker embedding $L$ is greater than an EEG window size $S$.  }

\label{Voice_encoder}

\subsection{Brain Encoder}
\label{Brain_encoder}
The brain encoder seeks to model the function of $B^{-1}(\cdot)$, as described in problem formulation, that derives a voice
signature from the EEG signals in a listening brain.
For tracking of a speaker's attention, we apply a sliding window of $S$ samples to the incoming EEG data stream, and detect the attended speaker window by window. It is assumed that the listener only attends to one speaker within an EEG window. 
To embrace a large receptive field in the EEG window, 
the brain encoder is built with dilated convolution network, as shown in Fig. \ref{Diagram} (c), to derive the voice signature.

For a given EEG decision window $\mathbf{e} \in \mathbb{R}^{S \times H}$, where $S$ represents the window size and $H$ represents the number of the EEG channels. A pointwise convolution block is used to combine the information across all EEG channels:
\begin{equation}
\mathbf{v}= 1 \times 1-\text{Conv}(\mathbf{e})
\end{equation}
where $\mathbf{v} \in \mathbb{R}^{S \times T}$ denotes the output latent representation and $T$ is the number of filters in the pointwise convolution block. 

Stacked 1-D dilated convolution network was proven effective in EEG signals characterization in various applications, such as stimulus-based AAD~\cite{accou2021modeling,puffay2022relating}, EEG-based seizure prediction~\cite{gao2022pediatric}, and EEG emotion recognition~\cite{li2021eeg}. In this paper, we are particularly interested in deriving the voice signature of the attended speaker from the EEG signals. We propose to adopt the stacked 1-D dilated convolution network for this purpose.
The latent representation $\mathbf{v}$ is then fed into $X$ stacked 1-D dilated convolutional blocks with increasing dilation factors to extract the EEG-derived voice signature. The dilation factors increase exponentially to ensure a sufficiently large temporal context window to take advantage of the long-range dependencies of the EEG signal. To minimize the number of parameters per receptive field size, we adopted the setting in \cite{accou2021modeling} that the dilated factors for the $X$ stacked 1-D dilated are 1, $P, \ldots, P^{X-1}$, where $P$ denotes the kernel size of these dilated convolutional blocks. A rectified linear unit (ReLU) function is adopted after each dilated convolutional block to ensure nonlinearity. For clarity, the dilated convolutional group is denoted as $D(\cdot)$. The EEG-derived voice
signature $\mathbf{b} \in \mathbb{R}^{S^{\prime} \times M}$ is:
\begin{equation}
\mathbf{b} = D(\mathbf{v})
\end{equation}
Where $M$ denotes the number of filters in the dilated convolutional blocks and $S^{\prime}$ denotes the length of the EEG-derived voice
signature $\mathbf{b}$. Noted the receptive size of $D(\cdot)$ is $P^X$, which should be smaller than the window size $S$.

\subsection{Attended Speaker Detector}
\label{Match_Estimator}
The attended speaker detector is a classifier that estimates the matching likelihood between a pair of audio-derived and EEG-derived voice signatures $[\widehat{\mathbf{s}}_j, \mathbf{b}]$.
As the two voice signatures are presented in the same middle hyperplane, we expect that they are close to each other if they are for the same speaker.
We therefore simply adopt the cosine similarity function
followed by a sigmoid function to measure the matching probability $p$ of a pair of audio-derived and the EEG-derived voice signatures
$[\widehat{\mathbf{s}}_j, \mathbf{b}]$: 
\begin{equation}
p = f\left(\widehat{\mathbf{s}}_j, \mathbf{b} \right) =\operatorname{Sigmoid}(\frac{\widehat{\mathbf{s}}_j \cdot \mathbf{b}}{\|\widehat{\mathbf{s}}_j\|\|\mathbf{b}\|})
\end{equation}
The speaker that has the highest matching probability, as formulated in Eq. (2), is detected as the attended speaker. 

\label{Loss_Function}
We adopt binary cross-entropy loss function to perform gradient descent for optimizing the E-ASD network:
\begin{equation}
\mathcal{L}=-y \log (p)-(1-y) \log (1-p)
\end{equation}
where $y \in \{0, 1\}$ indicates whether a pair of audio-derived and EEG-derived voice signatures matches, and $p$ is the predicted matching probability. 


\section{Experiments}
\label{sec:3}

We validate the effectiveness of the E-ASD on a well-known publicly available AAD database, which is referred to as the KUL database \cite{das_neetha_2020_3997352,vandecappelle2021eeg}.
We first train the E-ASD with two different publicly available speaker embedding models. We expected that E-ASD is a general approach that does not rely on a specific speaker embedding model.
We would also report the AAD performance in terms of
AAD accuracy, AUC, and EER with various network configurations.

\subsubsection{Speech-EEG Dataset}
The KUL database contains EEG data from 16 normal-hearing subjects when they selectively attend to one of two competing speakers.
Four 12-minute-long Dutch stories are narrated by 3 male Flemish speakers. In the four stories, silent segments longer than 500 ms were shortened to 500 ms.
Each story was then divided into two 6-minute-long sub-parts, yielding a total of 8 (4x2) different speech stimuli.
These 8 speech stimuli are presented to subjects in pairs during the experiments, resulting in four cocktail party scenarios. 
Furthermore, each cocktail party scenario is presented dichotically (one speaker per ear) or after head-related transfer function (HRTF) filtering to simulate speech with position information.
However, because we do not investigate the effect of direction information in this paper, these HRTF stimuli are not used in the experiment. 
Consequently, 8 trials $\times$ 6 minutes of EEG data were collected for each subject. 

In the KUL database, the two speech stimulus are provided in pairs with EEG data and attention labels. 
However, we aim to detect the voice of the attended speaker, which is less related to the speech content.
Therefore, the 8 speech stimuli are grouped by speakers, as shown in Table. \ref{Dataset_speaker}. The `partX\_trackX' are the stimuli names used in the KUL database.
Besides, an example of the customized 8 trials of Subject 1 is provided in Table. \ref{trial_example}. It should be noted that some trials have the same attended speaker but in different directions. For example, in both trials 1 and 2, Subject 1 is required to attend Speaker 1. However, Speaker 1's direction is right in Trial 1 but left in Trial 2. Therefore, with such a data structure, the E-ASD network can only learn to use voice information rather than direction information.

\input{dataset.tex}

\subsubsection{EEG Preprocessing}
Throughout the experiments, 64-channel EEG was recorded at a sampling rate of 8,192 Hz using a BioSemi ActiveTwo device. The EEG data were firstly re-referenced to the average response of all channels. As previous studies suggested that non-linear AAD decoders might benefit from broadband EEG information~\cite{de2020machine,vandecappelle2021eeg}, all the EEG data were bandpass-filtered between 1 and 32 Hz with a default finite impulse response (FIR) filter from MNE-Python and subsequently downsampled to 128 Hz in this study. Finally, the EEG channels were normalized to ensure zero mean and unit variance for each trial.

\subsubsection{Pretrained Speaker Encoder}
We leverage two different publicly available speaker embedding models: (1) ECAPA-TDNN embedder\footnote{\text{https://github.com/TaoRuijie/ECAPA-TDNN}} that generates 192-dimension speaker embedding; (2) ConvGRU embedder\footnote{\text{https://github.com/RF5/simple-speaker-embedding}} that generates 256-dimension speaker embedding. 
The ECAPA-TDNN model was trained on VoxCeleb2 dataset \cite{chung2018voxceleb2} with 5,994 speakers, while the ConvGRU embedder was trained on a combination of VCTK\cite{veaux2017cstr}, Librispeech\cite{panayotov2015librispeech}, and VoxCeleb2. 
Both embedders are widely adopted in speaker verification literature. 
The embedders take an utterance of arbitrary length, e.g., 5 to 10 seconds, as input, and generate a fixed size speaker embedding as output. In our experiments, we randomly select 10 speech segments with a length of 10 seconds from the stimuli `part1\_track1' as the enrollment utterances for Speaker1. Similarly, the enrollment utterances for Speaker2 and Speaker3 are selected from the stimulus `part3\_track1' and `part1\_track2' respectively. The pretrained speaker encoder derives the speaker embeddings $\mathbf{F}$ from these speech segments, thus the audio-derived voice signature $\widehat{\mathbf{s}}$ that serve as the pre-enrolled references.

\subsubsection{Training of E-ASD}

\textcolor{black}{We train the speaker encoder adaptor and brain encoder together by taking pre-enrolled utterances and an EEG window as the input. 
For clarity, each data point consists of three elements: [Speaker Embedding, EEG, Match Label]. For a Match-Mismatch structure, \cite{puffay2023relating} suggests that `the mismatched samples also appear as matched samples with other EEG segments' to ensure the model can generalize well. As shown in Table \ref{Dataset_speaker}, all three speakers have been attended or unattended in different trials, which satisfied the requirement. The E-ASD was trained in a subject-dependent manner, with the data for each subject randomly divided into three non-overlapped subsets, namely the training set (80\%), the validation set (10\%), and the test set (10\%). }

The hyperparameters of the E-ASD network are defined in Table~\ref{tab1}. We evaluate the network performance with different network configurations as discussed in Section \ref{parameter_search}. 
The networks are trained for 100 epochs on different decision window sizes of 0.5s and 1s with overlapping factors of 0.5 and 0.9. With a window size of 0.5s and an overlapping factor of 0.5, we obtain 18,384 Speech-EEG window pairs in the training set, 2,272 pairs in the validation set, and 2,272 pairs in the test set; With a window size of 0.5s and an overlapping factor of 0.9, we obtain 98,128, 12,112, and 12,112 pairs in the training, validation, and test sets, respectively.
With a window size of 1 second and an overlapping factor of 0.5, we obtain 9,168, 1,120, and 1,120 pairs in the training, validation, and test sets, respectively; With a window size of 1 second and an overlapping factor of 0.9, we obtain 48,976, 5,968, and 5,968 pairs in the training, validation, and test sets, respectively.

The Adam optimizer was adopted with the initial learning rate of $1 \times 10^{-3}$.
The learning rate is halved if the validation loss is not improved in 3 consecutive epochs.
The weights of convolution layers are initialized from a normal distribution (0 mean and 1 variance).

\input{AAD_parameter.tex}

\subsubsection{Evaluation}
At inference time, E-ASD takes an EEG window $\mathbf{e}$ as input, and detects whether $\mathbf{e}$ and the voice signature $\widehat{\mathbf{s}}$ forms a matching EEG-speech pair. We adopted three metrics to evaluate the performance of the E-ASD as follows, 
\begin{itemize}
    \item AAD Accuracy (ACC) is calculated as the percentage of the correctly detected trials over all trials.
    \item Area Under the receiver operating Characteristic (AUC) is a summary of the ROC curve that measures the overall detection performance. 
    \item Equal Error Rate (EER) reflects the fundamental of the E-ASD classifier, when the false acceptance rate (FAR) and false rejection rate (FRR) are equal.
\end{itemize}
A higher ACC or AUC value, or a lower EER value indicates a better performance.

\section{Results}
\label{sec:4}
We first examine how the pretrained speaker encoder behaves on the KUL dataset, then report the performance of E-ASD of different network configurations with three evaluation metrics.


\subsection{Performance of Pretrained Speaker Encoder}

\input{Embedding.tex}

To ensure that the pretrained speaker encoder works well on the KUL dataset, we evaluate the quality of resulting speaker embeddings $\mathbf{F}$. Specifically, for the 8 stimuli shown in Table \ref{Dataset_speaker}, we cut the first 10 speech segments with a length of 10 seconds from each stimulus and derive the speaker embedding.


We evaluate the cosine similarity between speaker embeddings of the same and different speakers for all speakers involved in KUL. We expect to see a high cosine similarity for the same speakers (e.g., `part1\_track1' and `part2\_track1') and a low cosine similarity otherwise. The results of ECAPA-TDNN embedder and ConvGRU embedder are shown in Table \ref{ECAPA-TDNN-result} and Table \ref{ConvGRU-result}, respectively.

We observe that, with the ConvGRU embedder, the similarity values are greater than 0.89 ($>0.89$) for the same speakers, and lower than 0.74 ($<0.74$) from different speakers. 
With the ECAPA-TDNN embedder, the similarity values are greater than 0.90 ($>0.90$) for the same speakers, but lower than 0.65 ($<0.65$) otherwise. The results suggest that the audio-derived voice signatures perform well as expected. %

\subsection{Performance of E-ASD Network}
\label{parameter_search}

\input{./AAD_results}

We report the ACC, AUC, and EER performance of E-ASD with several configurations in Table \ref{AAD_results}.
All the reported values are the averaged performance across all subjects and all folds. 

The experiments validated our hypothesis that it is possible to derive a voice signature from the elicited EEG signals, that is called EEG-derived voice signature, and to detect the attended speakers by comparing pre-enrolled audio-derived voice signature and EEG-derived voice signature. We also made the following observations.

\begin{enumerate}[(i)]
    \item EEG Window Size $(S, r)$: Increasing the overlapping factor or reasonably decreasing the decision window size increases the number of training samples and thus improves the performance. In previous stimulus-based approaches\cite{o2015attentional, biesmans2016auditory, crosse2016multivariate,wong2018comparison, de2020machine, thornton2022robust, xu2022decoding, de2018decoding, monesi2020lstm, accou2021modeling }, 
    the correlation between time-varying EEG signals and time-varying speech features is heavily relying on the temporal dependencies. Hence, a larger decision window size improves the temporal dependencies and usually results in better performance. In contrast, E-ASD detects the time-invariant voice signature from the time-varying EEG signals. The information contained in the audio-derived voice signature is not increased with a larger decision window. On the other hand, in a larger EEG decision window, more information is provided along with more noise. Therefore, unlike previous studies, E-ASD does not demonstrate an obvious performance benefit with larger window lengths.
    \item Speaker Embedder: ECAPA-TDNN embedder performs better than the ConvGRU embedder in E-ASD, possibly due to a better speaker embedding performance. As shown in Table. \ref{ECAPA-TDNN-result} and \ref{ConvGRU-result}, the speaker embeddings estimated by the ECAPA-TDNN embedder have larger distances between speakers.
    \item Number of filters in the $1 \times 1$ Conv block $T$: Increasing $T$ leads to better performance, as more features are combined from different EEG channels. 
    \item Number of filters in the 1-D Conv blocks $M$: Increasing $M$ increases the overcompleteness of the filters to extract the final audio-derived and EEG-derived voice signatures and therefore improves the performance. It should be noted that the model size and the computational complexity are also increased with $T$ and $M$. Therefore, there is a trade-off between the hyperparameters and performance. 
    \item Size of the receptive field $(P,X)$: The overall receptive field size in the dilated Conv blocks is $P^X$, which is limited by the EEG decision window size $S$. 
    Increasing the size of the receptive field leads to better performance, which shows the importance of modeling the temporal dependencies in the EEG signal. Specifically, the worst performance is achieved with $X=1$, in which the brain encoder is a single Conv1d layer without dilation. Besides, the scenarios $(X = 4, P=2)$ and $(X = 4, P=2)$ has the same receptive field sizes. Yet, the deeper network $(X = 4, P=2)$ demonstrates better performance, possibly due to the increased model capacity.
\end{enumerate}

\section{Discussion}

\label{sec:5}

\subsection{Comparison with Other AAD Tasks}
\label{V-A}
With KUL dataset, we validated the hypothesis that one can derive the voice signature of the attended speaker from the listener's EEG signals and detect the attended speaker. Unlike the prior AAD tasks where we detect the attention to speech content, speaker locus or direction of speech arrival~\cite{o2015attentional, biesmans2016auditory, crosse2016multivariate,wong2018comparison, de2020machine, thornton2022robust, xu2022decoding, de2018decoding, monesi2020lstm, accou2021modeling, geirnaert2020fast, vandecappelle2021eeg, su2022stanet, Deckers475673, kuruvila2021extracting, 9633231}, detecting the attended speaker in EEG signals is an unexplored novel task.  Therefore, it doesn't make sense to compare the accuracy with those of the previous works. Instead, we provide a comprehensive overview of various AAD tasks in Table~\ref{Comparison} to facilitate a functional comparison.

In locus-based methods~\cite{geirnaert2020fast, vandecappelle2021eeg, su2022stanet}, it is assumed that the locus of auditory attention is reflected in brain activities, that can be detected from the elicited EEG signals. However, in directional-stimulus-based~\cite{Deckers475673, kuruvila2021extracting, 9633231} or stimulus-based~\cite{o2015attentional, biesmans2016auditory, crosse2016multivariate,wong2018comparison, de2020machine, thornton2022robust, xu2022decoding, de2018decoding, monesi2020lstm, accou2021modeling } methods, it is assumed that the attended speech content is reflected in brain activities, thus, the auditory attention detection requires the availability of a spoken stimulus as the reference. The E-ASD task is different from the previous AAD studies in two ways.
First, we perform a hypothesis test as to whether a target speaker is being attended to, regardless of the content of speech stimulus or the speaker's locus.
Second, as the hypothesis test relies on the voice signature of the target speaker, it is required that the target speakers are pre-enrolled into the system. 

\input{./Qualitative_Comparison}

\subsection{Towards Real-Time Neurosteered Assistive Hearing} 
One promising application of AAD is the neurosteered hearing aids, which rely on the joint action of speech separation techniques and AAD \cite{ aroudi2020cognitive,das2020linear}.
However, speaker separation techniques typically necessitate knowledge of the number of speakers. Besides, hearing aids only present the attended speech signals to hearing-impaired users, separating all the speech signals seems unnecessary.
In a recent study, the Brain-inspired speech separation (BISS) model \cite{ceolini2020brain} reconstructs the attended envelope from the EEG signals and uses it as a reference cue to directly extract the attended speech signals. 
However, due to the noisy nature of the EEG signal, the use of the reconstructed `noisy envelope’ also introduces additional interference. Therefore, BISS sees a better performance with intracranial EEG
(iEEG) data. 

With E-ASD, we take a different path, that is to detect the match of audio-derived voice signature with the EEG-derived voice signature, that allows for pre-enrollment of audio-derived voice signature. We benefit from the research of speaker representation learning in speaker verification and speaker extraction~ \cite{xu2020spex}. In this way, a joint action for neurosteered speaker extraction between AAD and speaker extraction becomes straightforward. 

One critical implementation issue in neurosteered hearing aids is how to achieve real-time extraction during inference. 
The stimulus-based AAD methods \cite{aroudi2020cognitive,das2020linear, ceolini2020brain} rely on the speech stimulus as the reference, while the speaker extraction methods rely on the AAD output to identify the speech stimulus. Such interdependence complicates the implementation of the joint action between AAD and speaker extraction. As E-ASD relies on a pre-enrolled voice signature, such run-time interdependence is avoided.




Another factor of consideration in real-time implementation is the actual processing time during inference. With a decision window of 0.5 seconds and an overlapping factor of 0.9, E-ASD makes AAD decision every 50 milliseconds, which is much faster than the shifting rate of human attention in terms of seconds.

\subsection{Outlook}
In this paper, we present the feasibility of detecting the voice signature of the attended speaker from the EEG signals, which is a novel AAD task based on top-down attention. 
Despite much progress, we only test the E-ASD in a closed-set setup, that means the same speakers are involved in all training, validation, and testing sets. 
Indeed, most of the publicly available AAD datasets only consider the diversity of subjects, whereas the diversity of speakers is rarely considered. 

The E-ASD algorithm follows the implementation idea of reference-driven top-down attention, where the reference voice signature is pre-enrolled to the system. However, the stimulus-based algorithm follows the idea of bottom-up attention because the stimulus arrives at run-time without any pre-enrollment. In the human brain, speech perception is portrayed as a ``prediction machine'' where top-down expectations are constantly predicting bottom-up information~\cite{van2010brain}. 
For instance, when listening to the same speech content narrowed by different speakers, the collected EEG signals from the same subject might also be different. 
There are still open questions as to how we combine the top-down and bottom-up attentions in the AAD task. 

\section{Conclusion}

\label{sec:6}
\textcolor{black}{In a cocktail party scenario, EEG-based auditory attention detection is seen as an enabling technology for neurosteered assistive hearing. In this paper, we propose to detect the attended speaker from the listener's EEG signals. 
This study is motivated by the fact that humans have the ability to attend to a particular speaker easily in a complex multi-talker acoustic environment, and the hypothesis that such attention is reflected in the listener's EEG signals. 
We have proven that it is possible to detect the voice signature of the attended speaker from the listener's brain signal by successfully implementing the E-ASD framework. This study marks a non-trivial, but crucial step toward the EEG-controlled speech extraction.}


\bibliographystyle{IEEEtran}
\bibliography{mybib}

\end{document}

%% file: dataset.tex
\begin{table} [!b]
\centering
\caption{Stimulus-Speaker information}
\setlength{\tabcolsep}{3pt}
\renewcommand\arraystretch{1.7}
\begin{tabular}{p{60pt}<{\centering} p{60pt}<{\centering}  p{60pt}<{\centering}}
\hline\hline
Speaker1 & Speaker2 &Speaker 3
 \\
\hline
part1\_track1 & part3\_track1 & part1\_track2\\
part2\_track1 & part4\_track1 & part2\_track2\\
 & & part3\_track2 \\
 & & part4\_track2 \\
\hline\hline
\end{tabular}
\label{Dataset_speaker}
\end{table}

\begin{table} [!b]
\centering
\caption{An example of 8 trials for Subject1}
\setlength{\tabcolsep}{3pt}
\renewcommand\arraystretch{1.7}
\begin{tabular}{p{30pt}<{\centering}  p{50pt}<{\centering}  p{50pt}<{\centering}  p{60pt}<{\centering} }
\hline\hline
Trial &  Left Stimuli & Right Stimuli & Attended 
 \\
\hline
1 & Speaker3 & Speaker1 & Speaker1 (Right)
\\
2 & Speaker1 & Speaker3 & Speaker1 (Left)
\\
3 & Speaker3 & Speaker2 & Speaker2 (Right)
\\
4 & Speaker2 & Speaker3 & Speaker2 (Left)
\\
5 & Speaker1 & Speaker3 & Speaker3 (Right)
\\
6 & Speaker3 & Speaker1 & Speaker3 (Left)
\\
7 & Speaker2 & Speaker3 & Speaker3 (Right)
\\
8 & Speaker3 & Speaker2 & Speaker3 (Left)
\\
\hline\hline
\end{tabular}
\label{trial_example}
\end{table}

%% file: AAD_parameter.tex
\begin{table}[!t]
\centering
\caption{Hyperparameters of the E-ASD network evaluated on KUL dataset}
\label{tab1}
\setlength{\tabcolsep}{3pt}
\renewcommand\arraystretch{1.7}
\begin{tabular}{p{25pt}<{\centering} p{128pt} p{77pt}}
\hline\hline
Symbol & \makecell[c]{Description} & \makecell[c]{Setting}
 \\
\hline
\multirow{2}{*}{$L$} & \multirow{2}{*}{Dimension of speaker embedding} & ECAPA-TDNN: 192 \\ &~ & Conv-GRU: 256 \\
$H$ & Number of raw EEG channels & 64 \\
\multirow{2}{*}{$S$} & \multirow{2}{*}{EEG window size} & 64 for 0.5 seconds \\ &~ &128 for 1 second\\
$r$ & \makecell[l]{Overlapping factor of EEG window} & [0.5; 0.9]\\
$T$ & Number of filters in the $1\times1$ Conv block  & [2; 4; 8] \\
$M$& Number of filters of dilated 1-D Conv blocks & [4; 8; 16] \\
$X$& Number of dilated 1-D Conv blocks & [1; 2; 3; 4] \\
$P$& Kernel Size of dilated 1-D Conv blocks & [2; 3; 4] \\

\hline\hline
\end{tabular}
\label{tab1}
\end{table}

%% file: Embedding.tex
\begin{table}[!b]
    \centering
    \small
    \def\arraystretch{1.7}
    \setlength{\tabcolsep}{4.5pt}
    \caption{Cosine similarity of speaker embeddings $F$ between the same and different speakers (Spk1, Spk2, and Spk3) of ConvGRU embedder. The numbers in bold are for the same speakers.}
    \begin{tabular}{cc|cccccccc}
        \hline\hline
        \multirow{2}{*}{} & & \multicolumn{2}{c}{Spk1} & \multicolumn{2}{c}{Spk2} & \multicolumn{4}{c}{Spk3}\\
        \cline{3-10}
        ~ & ~& 1 & 2 & 1 & 2 & 1 & 2 & 3& 4\\
        \hline
        \hline
        \multirow{2}{*}{Spk1} & part1\_track1  & 1  \\
        ~ & part2\_track1 & \textbf{0.93} &  1\\
        \cline{1-2}
        \multirow{2}{*}{Spk2} &  part3\_track1  &  0.36 & 0.20 & 1\\
        ~ & part4\_track1 & 0.37 & 0.20  & \textbf{0.97} & 1\\
        \cline{1-2}
        \multirow{4}{*}{Spk3} & part1\_track2  & 0.59 & 0.45 & 0.73 & 0.73 & 1\\
        ~ & part2\_track2 & 0.51 & 0.35 & 0.74 & 0.74 & \textbf{0.95} & 1\\
        ~ & part3\_track2 & 0.56 & 0.42 & 0.69 & 0.69 & \textbf{0.95} & \textbf{0.93} & 1\\
        ~ & part4\_track2 & 0.59 & 0.48 & 0.65 & 0.65 & \textbf{0.95} & \textbf{0.89} & \textbf{0.91} & 1\\
        \hline\hline

    \end{tabular}
    \label{ConvGRU-result}
\end{table}

\begin{table}[!b]
    \centering
    \small
    \def\arraystretch{1.7}
    \setlength{\tabcolsep}{4.5pt}
    \caption{Cosine similarity of speaker embeddings $F$ between the same and different speakers (Spk1, Spk2, and Spk3) of ECAPA-TDNN embedder.}
            \begin{tabular}{cc|cccccccc}
        \hline\hline
        \multirow{2}{*}{} & & \multicolumn{2}{c}{Spk1} & \multicolumn{2}{c}{Spk2} & \multicolumn{4}{c}{Spk3}\\
        \cline{3-10}
        ~ & ~& 1 & 2 & 1 & 2 & 1 & 2 & 3& 4\\
        \hline
        \hline
        \multirow{2}{*}{Spk1} & part1\_track1  & 1  \\
        ~ & part2\_track1 & \textbf{0.96} &  1\\
        \cline{1-2}
        \multirow{2}{*}{Spk2} &  part3\_track1  &  0.47 & 0.40 & 1\\
        ~ & part4\_track1 & 0.47 & 0.40  & \textbf{0.98} & 1\\
        \cline{1-2}
        \multirow{4}{*}{Spk3} & part1\_track2  & 0.58 & 0.53 & 0.65 & 0.64 & 1\\
        ~ & part2\_track2 & 0.51 & 0.43 & 0.58 & 0.56 & \textbf{0.94} & 1\\
        ~ & part3\_track2 & 0.53 & 0.47 & 0.57 & 0.54 & \textbf{0.94} & \textbf{0.95} & 1\\
        ~ & part4\_track2 & 0.56 & 0.52 & 0.60 & 0.57 & \textbf{0.96} & \textbf{0.90} & \textbf{0.94} & 1\\
        \hline\hline

    \end{tabular}
    \label{ECAPA-TDNN-result}
\end{table}

%% file: AAD_results.tex
\begin{table} [!b]
\centering
\caption{Performance of E-ASD with various network configurations. (L = 256: ConvGRU, L = 192: ECAPA-TDNN)}
\vskip 0.05in
\renewcommand\arraystretch{1.7}
\begin{tabular}{ p{8pt}<{\centering}  p{8pt}<{\centering} p{8pt}<{\centering} p{4pt}<{\centering} p{4pt}<{\centering} p{4pt}<{\centering} p{4pt}<{\centering} p{32pt}<{\centering} | p{15pt}<{\centering} p{15pt}<{\centering} p{15pt}<{\centering}}
\hline\hline 
\multicolumn{8}{c|}{Network Parameters} & \multicolumn{3}{c}{Results}\\
\hline
\multirow{2}{*}{$L$} & \multirow{2}{*}{$S$} & \multirow{2}{*}{$r$} & \multirow{2}{*}{$T$} & \multirow{2}{*}{$M$} & \multirow{2}{*}{$X$} & \multirow{2}{*}{$P$}  & Receptive & ACC & AUC & EER\\
~ & ~ & ~ & ~ & ~ & ~ & ~ & field(s)  & (\%)$\uparrow$ & ~(\%)$\uparrow$ & ~(\%)$\downarrow$\\
\hline 
256 & 64 & 0.5 & 8 & 16& 3 & 3  & 0.211 & 97.18  & 98.51 & 3.22 \\
256 & 64 & 0.9 & 8 & 16& 3 & 3  & 0.211 & 99.34 & 99.72 & 0.80 \\
256 & 128 & 0.5 & 8& 16 & 3 & 3  & 0.211 & 96.36  & 97.74  & 5.41 \\
256 & 128 & 0.9 & 8 & 16 & 3 & 3  & 0.211 & 98.76 & 99.33 & 1.85 \\

\hline 
192 & 64 & 0.5 & 8 & 16& 3 & 3  & 0.211 &  98.08 & 99.20 & 2.22 \\
192& 64 & 0.9 & 8 & 16 & 3 & 3  & 0.211 &\textbf{99.78}  & \textbf{99.94} &\textbf{0.27} \\
192& 128 & 0.5 & 8  & 16 & 3 & 3 & 0.211 & 98.08 & 99.15 & 2.85\\
192& 128 & 0.9 & 8 & 16 & 3 & 3  & 0.211 &\textbf{99.78} & \textbf{99.94} & 0.34\\
\hline
192 & 64 & 0.9 &  2 &16  & 3 & 3  & 0.211 & 96.10  & 98.27 & 4.35 \\
192& 64 & 0.9 & 4 & 16 & 3 & 3  & 0.211 & 98.80  & 99.46  & 1.36 \\
192 & 64 & 0.9 & 8 & 4 & 3 & 3  & 0.211 & 97.28 & 98.82 & 3.12 \\
192& 64 & 0.9 & 8 & 8 & 3 & 3 & 0.211 & 99.23  & 99.74 & 0.93\\
\hline
192 & 64 & 0.9 & 8 & 16& 1 & 3  & 0.023 & 70.84  & 72.73 & 28.84 \\
192 & 64 & 0.9 & 8 & 16& 2 & 3  & 0.070 & 98.57  & 99.12 & 2.26  \\
192& 64 & 0.9 & 8 & 16& 2 & 2  & 0.031 & 94.26  & 95.59 & 6.47 \\
192& 64 & 0.9 & 8 & 16& 3 & 2  & 0.063 &  98.66 & 98.95 & 1.56\\
192& 64 & 0.9 & 8 & 16& 4 & 2  & 0.125 & 99.81 & 99.93 & \textbf{0.27} \\
192& 64 & 0.9 & 8 & 16& 2 & 4  & 0.125 & 99.32  & 99.72 &  1.06 \\
\hline\hline
\end{tabular}
\label{AAD_results}
\end{table}

%% file: Qualitative_Comparison.tex
\begin{table} [!b]
\centering
\caption{Functional comparison of several auditory attention detection tasks.} 
\setlength{\tabcolsep}{3pt}
\renewcommand\arraystretch{3}
\begin{tabular}{cccc}
\hline\hline
Approach & \makecell[c]{Reference signal} & Assumed availability &  \makecell[c]{Hypothesis test}
 \\
\hline
\makecell[c]{Locus-based \\ \cite{geirnaert2020fast, vandecappelle2021eeg, su2022stanet}} & N.A. & Speaker's locus & Locus \\ 
\hline
\makecell[c]{Directional- \\ stimulus-based \\ \cite{Deckers475673, kuruvila2021extracting, 9633231}} & \makecell[c]{Pre-ordered \\ stimulus} & \makecell[c]{Speech stimulus \\ Speaker's locus }& Speech content\\ 
\hline
\makecell[c]{Stimulus-based \\ \cite{o2015attentional, biesmans2016auditory, crosse2016multivariate,wong2018comparison, de2020machine, thornton2022robust, xu2022decoding, de2018decoding, monesi2020lstm, accou2021modeling }} & \makecell[c]{Stimulus} & \makecell[c]{Speech stimulus} & Speech content \\
\hline
\makecell[c]{E-ASD \\ (Proposed)}& \makecell[c]{Speaker \\ embedding} &\makecell[c]{Pre-enrolled speakers} & \makecell[c]{Speaker} \\

\hline\hline
\end{tabular}
\label{Comparison}
\end{table}